# Tailoring optical metamaterials to tune the atom-surface Casimir-Polder interaction

**Authors**
Eng Aik Chan,[1,3] Syed Abdullah Aljunid,[1] Giorgio Adamo[1], Athanasios Laliotis[2], Martial Ducloy*[1,2,3], David Wilkowski*[1,3,4,5]

* Corresponding authors' emails: martial.ducloy@univ-paris13.fr, david.wilkowski@ntu.edu.sg

**Affiliations**
[1] Centre for Disruptive Photonic Technologies, TPI, Nanyang Technological University, 637371 Singapore.
[2] Laboratoire de Physique des Lasers, UMR 7538 du CNRS, Université Paris13-Sorbonne-Paris-Cité   F-93430 Villetaneuse, France.
[3] School of Physical and Mathematical Sciences, Nanyang Technological University, 637371, Singapore.
[4] Centre for Quantum Technologies, National University of Singapore, 117543 Singapore.
[5] MajuLab, CNRS-UNS-NUS-NTU, Université Côte d'Azur, International Joint Research Unit UMI 3654, Singapore.

**Abstract**

Metamaterials are fascinating tools that can structure not only surface plasmons and electromagnetic waves but also electromagnetic vacuum fluctuations.  The possibility of shaping the quantum vacuum is a powerful concept that ultimately allows engineering the interaction between macroscopic surfaces and quantum emitters such as atoms, molecules or quantum dots. The long-range atom-surface interaction, known as Casimir-Polder interaction, is of fundamental importance in quantum electrodynamics but also attracts a significant interest for platforms that interface atoms with nanophotonic devices.  Here we perform a spectroscopic selective reflection measurement of the Casimir-Polder interaction between a $Cs(6P_{3/2})$ atom and a nanostructured metallic planar metamaterial. We show that by engineering the near-field plasmonic resonances of the metamaterial, we can successfully tune the Casimir-Polder interaction, demonstrating both a strong enhancement and reduction with respect to its non-resonant value. We also show an enhancement of the atomic spontaneous emission rate due to its coupling with the evanescent modes of the nanostructure. Probing excited state atoms next to nontrivial tailored surfaces is a rigorous test of quantum electrodynamics. Engineering Casimir-Polder interactions represents a significant step towards atom trapping in the extreme near field, possibly without the use of external fields.

# Introduction

Material boundaries modify the surrounding quantum vacuum giving rise to interactions between classical macroscopic surfaces (Casimir effect), or between a surface and a quantum polarizable object (Casimir-Polder interaction) [1]. The surface, inducing a shift of atomic energy levels, was investigated experimentally with thermal vapors [2], [3], [4], atomic beams [5], [6] and cold atoms [7],[8] [9]. The dispersive energy shift goes hand in hand with a dissipative change of the atomic radiative properties, allowing for the tuning of spontaneous emission rates of atoms next to dielectric surfaces. For this purpose, placing atoms next to tailored nanostructures is becoming a new challenge in the field of nanophotonics. Hybridization of atomic and nanophotonic systems paves the way to novel quantum devices due to strong atom-light coupling with micro-cavities [10] or due to collective effects arising from strong confinement [11], [12]. The most common hybrid platforms include tapered nanofibers [13], hollow core fibers [14], or photonic bandgap waveguides [15].

The Casimir-Polder force has been largely perceived as an obstacle for placing atoms close to surfaces. Nevertheless, ambitious proposals emerge that suggest the possibility of utilizing atom-surface interactions to achieve tight trapping at record distances from surfaces [16] and in particular photonic bandgap waveguides [17], [18]. The tantalizing possibility of subwavelength atom trapping is made more difficult due to the predominantly attractive nature of the Casimir-Polder interaction that does not allow stable trapping potentials in all directions. Experimental demonstration of an efficient tuning of the atom-surface interaction, particularly between attraction and repulsion represents therefore a milestone in the field of hybrid systems.

Among the various approaches to modify the Casimir-Polder interaction, one relies on the resonant coupling between excited atoms and surface resonances. Thus, experiments with Cesium atoms in high lying excited states next to a sapphire surface have demonstrated resonant Casimir-Polder repulsion [19], [20] or an exaltation of the Casimir-Polder attraction with temperature [21]. The key parameter here, which governs both strength and sign of the Casimir-Polder interaction, is the relative detuning of the plasmonic resonance frequency compared to the predominant atomic dipole coupling (transition). Active engineering of the atom-surface interaction is therefore severely limited by the selection of dielectrics that are available in nature. The possibility of exploring material birefringence as a means of tuning the polariton resonances has been proposed [22], but the theoretical tunability is restricted and yet to be demonstrated experimentally.

An alternative solution would be to use nanostructured periodic planar metamaterials that allow a broad tunability of plasmonic surface resonances across the visible and near infrared spectrum. This spectral domain, where plasmonic or polaritonic resonances are usually scarce in bulk material, is of particular interest because it matches with low-lying transitions of alkali atoms, routinely used in laser cooling or atomic spectroscopy experiments. Moreover, thin metallic planar metamaterials have enhanced light transmission at the plasmon resonance. It thus becomes possible to perform reflection spectroscopy on a vapor/metamaterial interface. Previous experiments on Cesium vapor demonstrated a modification of the surface reflectivity, however, analysis of the Doppler broadened spectra did not provide quantitative information on the frequency shift or the atomic lifetime modification at the proximity of the metamaterials [23], [24].

Here, we report on high-resolution frequency-modulated selective reflection (SR) spectroscopy of Cs($6P_{3/2}$) atoms in vicinity of a wide range of metallic planar metamaterials. The theoretical treatment of SR on flat dielectric surfaces [25] is here further developed, allowing us to measure the dispersive Casimir-Polder shift of the $6S_{1/2}\leftrightarrow6P_{3/2}$, Cesium D2 transition as well as a dissipative modification of the spontaneous emission rate [26]. For an adequately chosen metamaterial, the frequency shift on the Cesium transition, induced by the Casimir-Polder interaction, can be almost suppressed. A quantum electrodynamics (QED) calculation of the fully retarded Casimir-Polder interaction is shown to reproduce the basic features of our experimental data.

## Results

The system under investigation, similar to the one used by Aljunid *et al.* [24], is depicted in Fig.1a. A Cesium vapor at T=80°C is introduced into a vacuum chamber. The density of the atoms is around N=$10^{17}$ m$^{-3}$ whereas the thermal velocity is $\bar{u}$=150ms$^{-1}$. On one of the dielectric viewports of the vacuum chamber, 10 different metamaterials are engraved on a 50nm thick silver layer (see Fig.1b). Each metamaterial consists of an area of 200 µm x 200 µm square area containing arrays of 70-nm-wide nanoslits of varying length from 170 to 240 nm. The unit cells are squares with sizes from w = 380 to 520 nm long, depending on the length of the slit. We achieve plasmonic resonances, characterized by their central position, covering a wavelength range from 670 to 980nm. Each plasmon resonance has a typical width of 60nm (see an example in Fig.1c). The atoms/metamaterial hybrid system is excited and probed on the $6S_{1/2}$↔$6P_{3/2}$ cesium D2 transition at 852nm, using an SR optical setup at normal incidence [2], [27], [28].

The radiation is produced by an external cavity diode laser (ECDL). Part of the light is sent to a saturated absorption spectroscopy setup used as a frequency reference. The main beam goes through an electro-optic modulator to modulate its phase at $\omega_m / 2\pi$= 9MHz with a modulation index of β=0.19 and is then shaped by a square mask to optimize its overlap with the metamaterial. The intensity of the laser is maintained below saturation, typically I = 1mW/cm$^2$. The reflected light beam is collected on a fast and sensitive amplified photodetector. The signal at 9 MHz is demodulated using a lock-in amplifier, directly providing the in-phase and in-quadrature component of the reflection signal, which is purely Doppler-free and analyzed in the following.

First, we analyze the reflectance and transmittance of the metamaterials at 852 nm using the ECDL. To do so, we detuned the laser away from the cesium resonance such that it is not coupled anymore to the atomic vapor whereas the coupling to the plasmon remains unchanged. The results are summarized in Fig.2. Here, each pair of reflectance/transmittance data corresponds to a given metamaterial, which is characterized by its center resonance frequency, $\lambda_p$, reported on the x axis of the graph. The resulting curves show a smooth resonance-like behavior, indicating that all the metamaterials have a similar reflectance and transmittance spectrum (see also the study of Aljunid *et al.* [24]). For a more quantitative analysis, we perform a Finite Difference Frequency Domain (FDFD) simulation of the metamaterials. We extract the expected far-field reflectance and transmittance intensities of the metamaterials. The FDFD results are in good quantitative agreement with the experimental data. However, we perform a global adjustment of the contrast of the reflectance and transmittance of the FDFD results by a factor of 0.7 and 0.5 respectively. This adjustment accounts for frequency-independent optical losses, which are not taken into account in the FDFD simulation. Contributions may come from photon scattering at the metamaterials, imperfect size matching of the beam to the metamaterial, or possible long-range inhomogeneity of the metamaterial geometry.

We now tune the laser onto the F = 4↔F' = 5 hyperfine transition scanning over a frequency range of 100MHz = ~ 20Γ / 2π, where Γ = 5.2MHz, is the bare frequency width of the atomic transition. The demodulated signals at 9MHz of the beam reflectance correspond to the red points on Fig.3. We observe a strong modification of the Doppler-free spectrum due to the presence of the metamaterials with respect to a plain dielectric window (see blue points on Fig.3). The hyperfine structure of the D2 line is spectrally resolved with our SR experiment, because the frequency spacing between hyperfine components (200MHz between F'=3 and F'=4 and 250MHz between F'=4 and F'=5) are much larger than the width of the observed spectra (~10 to 20MHz). Also, the plasmonic resonance is much broader than the hyperfine splitting therefore its effects are identical on all hyperfine components. Similar SR spectra can be obtained for the F=4↔F'=3 and 4 transitions albeit with a smaller amplitude due to weaker transition probabilities.

To further analyze the experimental data, we model the metamaterial by a spatially homogeneous bulk material having the same thickness and far-field reflectance/transmittance at normal incidence. This mean-field approximation holds for the thermal vapor at the vicinity of the metamaterial because the large spatial variation of the plasmon (see example in Fig.1d) is well smoothed by the finite response time of the atomic coherence (see Supplementary Materials for more details).

In SR spectroscopy with a homogeneous material interface, the vapor can be characterized by an effective electric susceptibility, $\chi$. Its frequency derivative is given, in the Doppler limit ($k\,\bar{u} \gg \Gamma$), by [25]:

$$\frac{d\chi(\omega)}{d\omega} = \frac{2Nk\mu^2}{\sqrt{\pi}\,\bar{u}\,\epsilon_0\hbar} \int_0^\infty dz \int_0^\infty dz' \frac{(z-z')e^{ik(z+z')}}{\mathcal{L}_0(\omega,z) - \mathcal{L}_0(\omega,z')} \tag{1}$$

$\mathcal{L}(\omega,z)/z = \frac{\Gamma}{2} - i\left(\omega - \omega_o + \frac{\Delta C_3 z^{-3}}{2}\right)$ is the Lorentzian lineshape of the bare atomic resonance, corrected by the $z^{-3}$ Casimir-Polder frequency shift in the nonretarded regime. Here, $\omega_o$ is the bare atomic resonance, $k$ is the wavenumber of the laser beam and $\mu$ is the two-level atomic dipole moment. Spectroscopic experiments are sensitive to the energy difference between levels. We therefore denote $\Delta C_3$ the difference of van der Waals coefficients between lower and upper state. In our analysis, $\Delta C_3$ is considered to be complex. As can be seen from the equations above, its real and imaginary parts denote a distance-dependent shift and linewidth respectively. The metamaterial/vapor interface is located at z = 0. We note that the phase factor, $e^{ik(z+z')}$ rapidly averages the effective susceptibility to zero for $z, z' \gg k^{-1}$. As an important consequence, only the atoms located in a layer of thickness $k^{-1}$ contribute to the SR signal. Moreover, as can be seen in Eqn.1 and already discussed above, the signal is Doppler-free. We note as well that, because $\chi \ll 1$, the index of refraction of the atomic vapor reads as $n = 1 + \chi/2$. Under this approximation, the complex reflection coefficient of the electric field for the dielectric/metamaterial/vapor system can be linearized as:

$$r = r_o + \rho\chi \tag{2}$$

where $r_o$ and $\rho$ depend only on the indices of refraction of the metamaterial $n'$ and of the dielectric substrate $n_d$ (see Materials and Methods). In the Doppler limit, $\chi(\omega)$ is obtained by integration of Eqn. 1. In the weak-modulation limit, i.e. $\beta \ll 1$ we find that the demodulated signal has the following expression for the in-phase signal [25]:

$$V_p(\omega) = V_0 Re\{r_0^*\rho[\chi(\omega + \omega_m) - \chi(\omega - \omega_m)]\} \tag{3}$$

and for the in-quadrature signal,

$$V_q(\omega) = V_0 Im\{r_0^*\rho[\chi(\omega + \omega_m) + \chi(\omega - \omega_m) - 2\chi(\omega)]\} \tag{4}$$

$V_0$ is a factor of proportionality. Within the mean-field approximation, the complex factor $r_0^*\rho$ can be evaluated analytically (see Materials and Methods for more details). It depends on $n'$, $n_d$ and the metamaterial thickness. By mixing the real and imaginary part of the susceptibility, the factor $r_0^*\rho$ gives the main contribution of the modification of the atomic resonance line shape induced by the plasmon resonance observed in Fig.3. Also, the product of the susceptibility with complex value $r_0^*\rho$ in Eqn.3 and Eqn.4 leads to Fano-like resonance of the atoms/metamaterial hybrid system as shown by some of us in the study of Aljunid *et al.* [24] and by Stern and coauthors using a Kretschman geometry [23]. In addition to the Fano-like resonance, the Casimir-Polder interaction induces an additional contribution to the SR signal that we are now aiming to reveal and discuss.

Using Eqns.1 to 4, we perform a fit of the SR signals for the different metamaterials including the bare dielectric substrate. The fitting parameters are the atomic resonance linewidth $\Gamma$, the complex value of $\Delta C_3$ and a common $V_0$ value. The effective index of refraction of the metamaterial n', thus the factor $r_0^*\rho$ is extracted from the FDFD simulation [29], [30]. The results of the fitting procedure correspond to the black curves in Fig.3. We observed an agreement with the experimental data. The volume atomic resonance width (away from the surface) is found to be $\Gamma / 2\pi = 10(3)$ MHz, that is, slightly larger than the bare linewidth of 5.2MHz. This increasing of the linewidth, encountered as well on the dielectric interface, has been also reported in similar studies [3]. It can be due to residual collisional broadening. An imaginary part for the $\Delta C_3$ has to be considered. To illustrate this point, we perform another fit, setting Im[$\Delta C_3$] = 0. Under this constraint, we observe that the convergence of the fitting procedure is not satisfactory (see residue comparison in Fig.3 for $\lambda_p = 858$nm).

The complex van der Waals coefficients obtained from the fits are shown in Fig.4a-b. The real part of the $\Delta C_3$ coefficient (Fig.4a), displays a dispersive type of resonant behavior centered at $\lambda_p \sim 840$nm. At the blue side of the resonance, we observe a significant increase of the interaction, with respect to its off resonant value $\sim 5$kHz μm$^3$, followed by a sharp decrease that leads to a nearly vanishing value of the interaction at the red side of the resonance. The resonance width is in agreement with the plasmon linewidth of 60nm, confirming the plasmonic origin of the modification of the $\Delta C_3$ values. The presence and evolution of the imaginary (dissipative) part for the van der Waals coefficients, shown in Fig.4b, corresponds to a decreasing of the atomic lifetime associated with enhancement of the vacuum mode density at the plasmonic resonance. This enhanced emission rate can also be understood as an increasing of the Purcell factor as observed, for example, with ultracold gas [31] or nano-antennas [32]. The enhanced radiation emission of the atom/metamaterial system can finally either be coupled to electromagnetic propagating modes or be lost due to Ohmic losses in the metal. The selective reflection technique does not distinguish between those two cases.

**Discussion**

We compare our experimental measurements, with the QED theory of atom-surface interactions. A complete analysis of the resonant Casimir-Polder interaction depends on knowledge of the dielectric properties of the metamaterial for both real and imaginary frequencies. For this purpose, we fit the dielectric constant ε, extracted by FDFD simulations, to an analytical model that accounts for the resonances of the metamaterial as well as the surface plasmon resonance of silver itself. We then calculate the difference of the Casimir-Polder frequency shift between Cs(6P$_{3/2}$) and Cs(6S$_{1/2}$), which is the experimentally measured quantity in SR spectroscopy. Here, we take into account both non-resonant and resonant components. The nonresonant term does not depend on the position of the plasmon resonance and is mostly governed by the metallic response over the entire frequency spectrum [33], [34]. The influence of the plasmon resonance is contained in the resonant contribution [34], [35], which in our case is only relevant for the excited state Cs($6P_{3/2}$) interaction potential. It concerns the resonant photons on the $6S_{1/2} \leftrightarrow 6P_{3/2}$ transition and it is mainly at the origin of the observed Casimir-Polder resonant behavior (see Supplementary Materials). The results of the non-retarded model are shown as dashed lines in Fig.4c-d. The theoretical predictions exhibit a resonant behavior similar to the experimental findings, showing that our model captures the essence of the physical mechanism behind the tuning of the atom-surface interaction. However, the amplitude of the resonance is smaller than the experimentally measured one.

In our experiments, the plasmonic resonances coincide with the probing SR wavelength. This suggests that we are not in a pure near-field regime and that retardation effects cannot be neglected [36]. In the retarded regime the dephasing of the oscillating dipole with respect to its image causes the resonant term of the excited state to oscillate with a period of $\lambda/2$ (atomic transition wavelength) [33], [34]. The phase and amplitude of these oscillations depends on the surface reflection coefficient [36] which varies significantly with the plasmonic resonance of the metamaterial. For this reason, retardation can affect the amplitude and the shape of the $\Delta C_3$ resonance (see also Supplementary Materials). To strengthen our analysis further, we calculate an 'effective' van der Waals coefficient as a function of distance, defined as $C_3(z) = \frac{\delta E(z)z^3}{\hbar} + i\frac{\delta \Gamma(z) z^3}{2\hbar}$, where $\delta E(z)$ is the Casimir-Polder shift and $\delta\Gamma(z)/2$ the linewidth of a given atomic level. We find that they both deviate significantly from the near field (van der Waals approximation) even at nanometric distances away from the metamaterial surface. Analysis of the SR line shapes reveals that the relevant distance range is 70-100nm, in agreement with a characteristic probing depth $\sim \lambda/2\pi$. In Fig.4c-d we compare our experimental results with the real and imaginary parts of an effective $\Delta C_3(z)$ coefficient for distances between 70nm and 100nm, where $\Delta C_3(z)$ is the difference between the effective van der Waals coefficients of excited ($6P_{3/2}$) and ground states ($6S_{1/2}$). Retardation effects enhance the amplitude of the $\Delta C_3$ resonance but fail to exactly reproduce the experimental data. It should be noted that our calculations were performed for a semi-infinite effective bulk material. However, considering a finite layer of 50nm has a minor impact on the $\Delta C_3$ values (see Supplementary Materials for more details). We also mention that the real and imaginary parts of the $\Delta C_3$ coefficient in the case of pure silver are ~4.5 and ~0.5kHz $\mu m^3$ respectively. This is in agreement with the experimentally measured off-resonant values of the coefficient, confirming the validity of our data analysis. We finally note that in our mean field analysis, we have considered a homogeneous metamaterial characterized by one effective dielectric constant. A future more detailed analysis would have to take the anisotropy of the metamaterial into consideration. For this purpose, the theoretical framework for anisotropic interactions described by Gorza *et al.* [22] has to be expanded to include the effects of retardation.

To conclude, we use planar metamaterials as a testbed to control the Casimir-Polder frequency shift of the D2 line of Cs atoms. When the plasmon resonance almost coincides with the atomic resonance, the Casimir-Polder frequency shift nearly vanishes. Across the different metamaterials, the Casimir-Polder interaction is characterized by the usual real coefficient leading to a dispersive resonance, as well as a non-zero on-resonance dissipative response.

Finally, we mention that gratings have been previously used in Casimir [37], [38] as well as non-resonant Casimir-Polder experiments with Bose-condensed ground-state Rb atoms [39]. In these cases, the experiments have strenuously tested various theoretical models/approximations for calculating vacuum forces with non-trivial surface geometries, demonstrating the non-additivity of the Casimir-Polder interaction at short distances. Our measurements are different since we explore resonant Casimir-Polder interactions with excited state atoms next to a silver metamaterial.

It might be interesting to note that our results show that nano-patterning the silver surface, typically removing 7% of material, can in some cases increase the Casimir-Polder interaction by more than 100% with respect to pure silver. This is a powerful illustration of Casimir-Polder non-additivity. Our theoretical analysis, based on the mean field approximation, reproduces the main characteristics of the Casimir-Polder resonance albeit with a smaller amplitude. It would be worth comparing our experimental results with numerical models that fully account for the shape of the metamaterial.

**Materials and Methods**

**Sample fabrication**
Our experiments were conducted in a vacuum chamber with metamaterial samples fabricated on one of the optical access windows of the chamber [24]. A 3° quartz wedge window is used to support a 50-nm layer of silver deposited by thermal evaporation and protected by an 8-nm layer of $SiO_2$. The metamaterials were then fabricated by focused ion beam milling. Each metamaterial array was 200 µm × 200 µm in size. We also fabricated a window in the silver film for reference measurements. The chamber base pressure is $10^{-8}$ mbar, whereas the cesium vapor pressure is maintained at a temperature of 80°C corresponding to vapor pressure of $P = 10^{-4}$ mbar.

**Derivation of the $r_0^* \rho$ parameter**
The reflection coefficient in the mean-field approximation is given by [40] :

$$r = \frac{(n_d-n')(n'+n)+(n_d+n')(n'-n)e^{\frac{2in'l\omega}{c}}}{(n_d-n')(n'+n)-(n_d+n')(n'-n)e^{\frac{2in'l\omega}{c}}} \tag{5}$$

where $n_d$, $n'$ and $n$ are respectively the index of refraction of the dielectric substrate, of the effective bulk material, (describing the metamaterial) and of the atomic vapor. $l$ is the effective bulk material thickness. The index of refraction of the atomic vapor is given by $n = 1 + \chi/2$ where the atomic susceptibility $\chi$. Because $\rho = \frac{\partial r}{\partial \chi}|_{\chi=1}$, the complex coefficient $r_0^* \rho$ is found to be:

$$r_0^* \rho = \frac{4|r_o|^2 n_d n'^2 e^{\frac{2in'l\omega}{c}}}{(1+n')^2(n'^2-n_d^2)+2(1-n'^2)(n_d^2+n'^2)e^{\frac{2in'l\omega}{c}}+(1+n')^2(n'^2-n_d^2)e^{\frac{4in'l\omega}{c}}} \tag{6}$$

**Calculation of the Casimir-Polder interaction**
Here, we calculated the Casimir-Polder interaction between an excited state atom and a metamaterial using the perturbation approach first described by Wylie and Sipe [33], [34], also followed in numerous other works. In the framework of the mean field approximation, metamaterials were considered uniform and isotropic materials with an effective dielectric constant extracted by FDFD simulations for a wavelength range between 500nm to 1500nm. Casimir-Polder calculations rely on the knowledge of the dielectric function at the entire frequency range and for this purpose, we fit the FDFD data using the following analytic function:

$$\epsilon(\omega) = 1 - \frac{\omega_p^2}{\omega^2+i\gamma_p\omega} + \sum_j \frac{f_j \omega_j^2}{(\omega_j^2-\omega^2)-i\gamma_j\omega} \tag{7}$$

The first part of the equation is a Drude model that accounts for the bulk properties of silver. The plasmon frequency $\omega_p$ and dissipation $\gamma_p$ are common for all metamaterials. The second part, models the metamaterial resonances. The amplitude ($f_j$), frequency ($\omega_j$) and dissipation ($\gamma_j$) of each resonance are smoothly varying functions of the plasmonic resonance ($\lambda_p$) of the metamaterial. By interpolating these parameters we can calculate the Casimir-Polder interaction for a continuous range of metamaterials (see Fig.4c,d). We stress in this study that our calculation does not critically depend on the analytic model used to fit the dielectric constant. This is because the physics is mostly contained on the resonant part of the interaction that depends on $Re[\frac{\epsilon(\omega_o)-1}{\epsilon(\omega_o)+1}]$, where $\omega_o$ is the atomic resonance frequency. At zero temperature (T=0) the ground state potential as a function of distance z, writes:

$$\delta E(z) = -\frac{1}{\pi}\sum_n \omega_{gn}\mu_\alpha^{gn}\mu_\beta^{ng} \int_0^\infty d\xi \frac{G_{\alpha\beta}(z,i\xi)}{\xi^2+\omega_{gn}^2} \tag{8}$$

The excited state potential is as follows

$$\delta E(z) = -\frac{1}{\pi}\sum_n \omega_{en}\mu_\alpha^{en}\mu_\beta^{ne} \int_0^\infty d\xi \frac{G_{\alpha\beta}(z,i\xi)}{\xi^2+\omega_{en}^2} - \sum_n \mu_\alpha^{gn}\mu_\beta^{ng} Re[G_{\alpha\beta}(z,|\omega_{en}|)](1-\Theta(\omega_{en})) \tag{9}$$

whereas the distance dependent atomic level linewidth is given by

$$\delta\Gamma(z) = 2\sum_n \mu_\alpha^{gn}\mu_\beta^{ng} Im[G_{\alpha\beta}(z,|\omega_{en}|)](1-\Theta(\omega_{en})) \tag{10}$$

Here the summation has to be made on all possible dipole couplings terms. The transition frequencies $\omega_{gn}$ or $\omega_{en}$ are considered positive for upward couplings and negative for downward couplings. $\Theta(\omega_{en})$ is the Heaviside function, $\xi$ is an integration variable, $\mu_\alpha^{gn}, \mu_\beta^{ng}, \mu_\alpha^{en}, \mu_\beta^{ne}$ are the dipole moment matrix elements and $G_{\alpha\beta}(z,i\xi)$ is the linear susceptibility function defined by Wylie and Sipe [33], [34], and by Laliotis and Ducloy [36]. We used the Einstein notation, implying a summation over the index variables α and β that denote the Cartesian coordinate components. In the near field, the linear susceptibility is a diagonal matrix whose elements are proportional to $\frac{1}{(2z)^3}\frac{\epsilon(\omega)-1}{\epsilon(\omega)+1}$.

**Supplementary Materials**

Supplementary materials are appended below in the same file.

Stationary surface plasmon waves
Mean field approximation for the atomic vapor response
Resonant photon contribution to the Casimir-Polder interaction
Retardation effects on selective reflection
Finite thickness of the metamaterial
Fig. S1. Electric field simulations.
Fig. S2. Selective Reflection spectra.
Fig. S3. Resonant contribution to the Casimir-Polder interaction.
Fig. S4. Retardation effects on selective reflection.
Fig. S5. Finite thickness of the metamaterial.

**Acknowledgments**

The work was supported by the Singapore Ministry of Education Academic Research Fund Tier (Grant No. MOE2011-T3-1-005). A.L. thanks the UMI Majulab and the Centre for Quantum Technologies for supporting his trip to Singapore. We wish to thank Nikolay I. Zheludev and Emmanuel Lassalle for fruitful discussions.

**Figures and Tables**

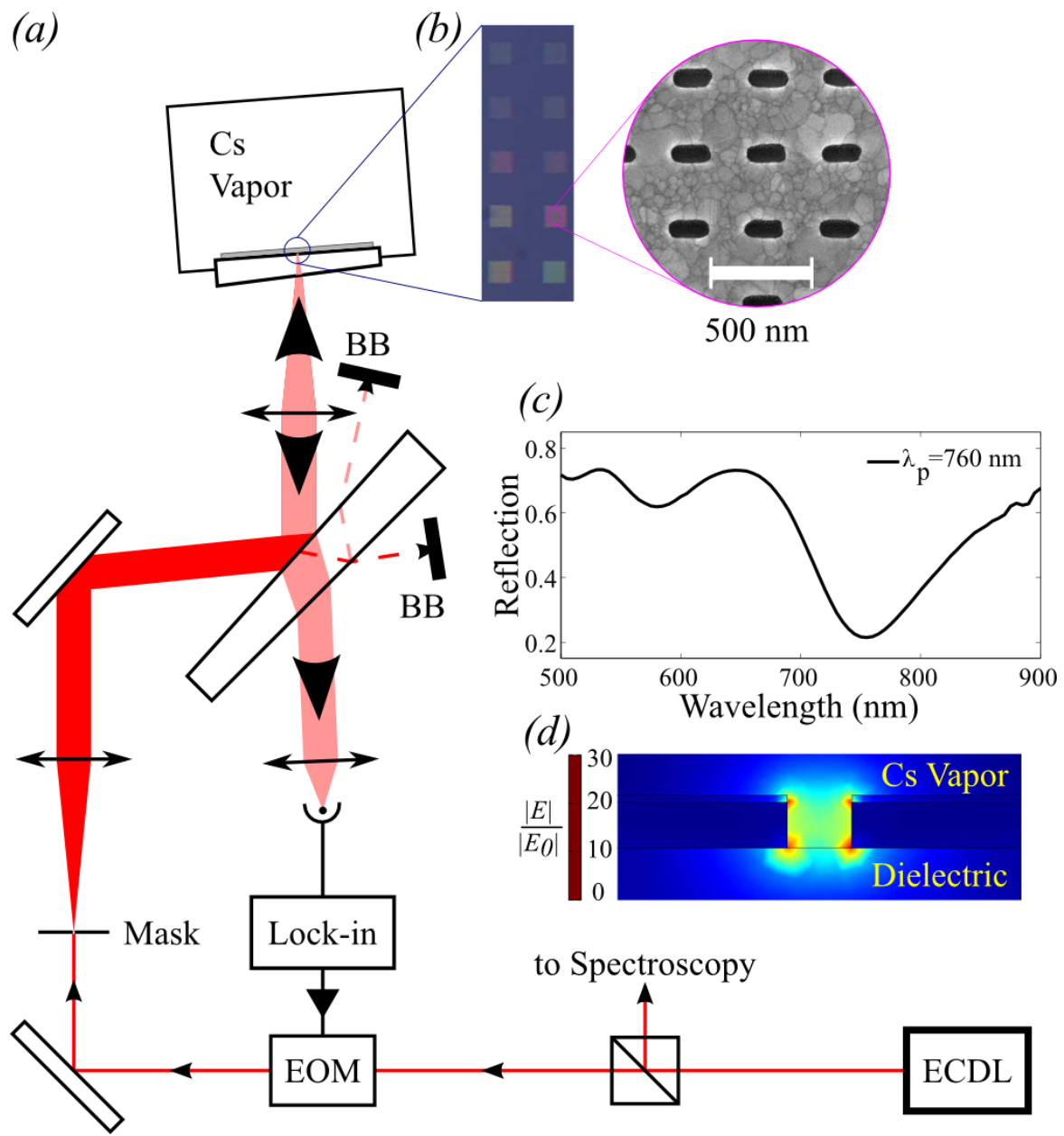

**Fig. 1. Schematic of the experimental set-up.** (a) Experimental set-up. (b) Real color back-illuminated images of the ten metamaterials. The zoom corresponds to an SEM

image. (c) A typical reflectance curve of a metamaterial showing a main plasmon resonance at λ=760nm. (d) Cut along one nanoslit. The false colors represent the electric field magnitude, normalized by the amplitude of the incident field, as obtained by a FDFD simulation. EOM, electro-optic modulator; BB, beam block.

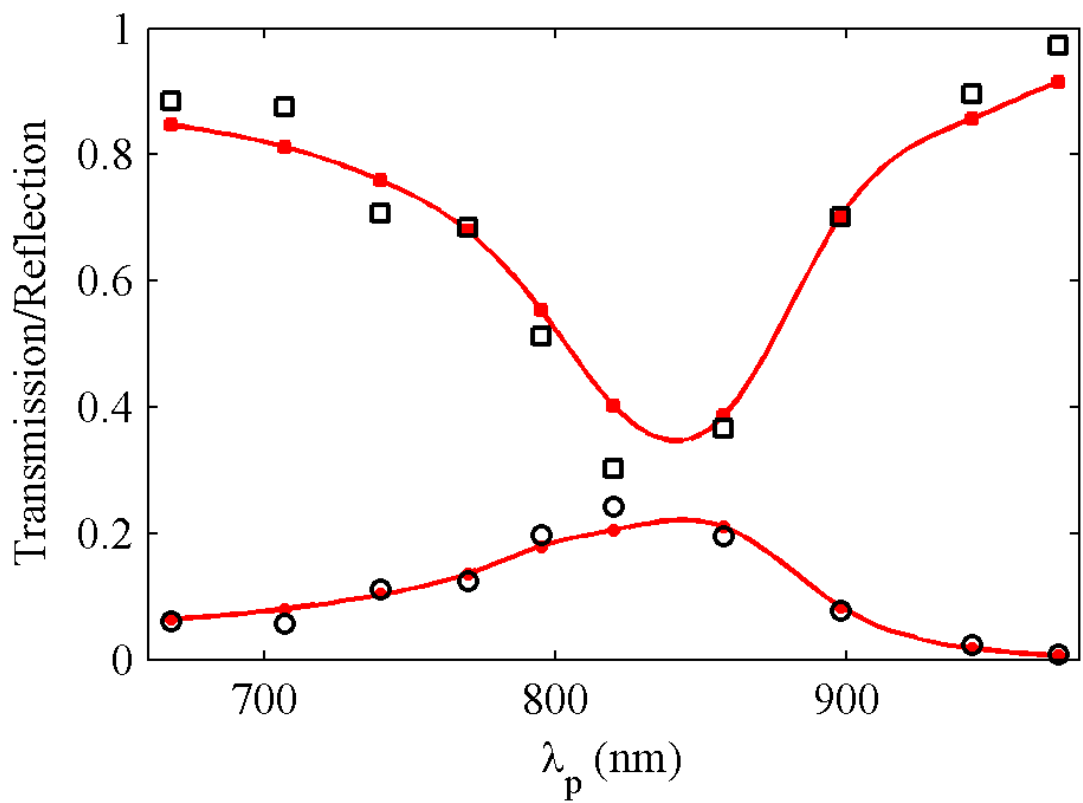

**Fig. 2. Optical characterization of the metamaterials.** Experimental reflection (open squares) and transmission (open circles) of the 10 metamaterials measured with the 852-nm laser. The x-axis corresponds to $\lambda_p$, the position of the plasmon resonance of each metamaterial. The solid red squares (circles) corresponds to the reflection (transmission) obtained by a FDFD numerical simulation. The lines, connecting FDFD results, are guides for the eye.

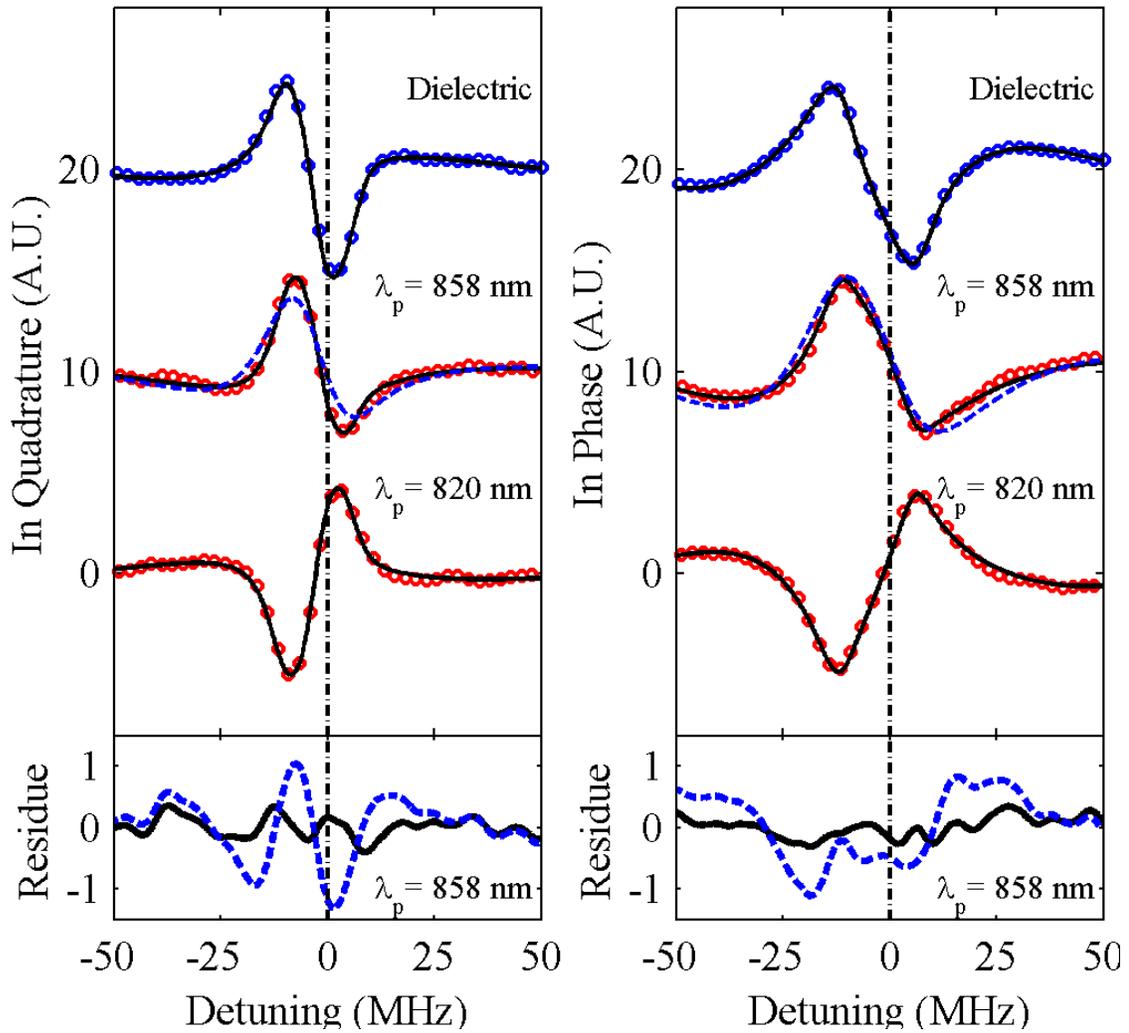

**Fig. 3. Selective reflection spectra.** In-phase and in-quadrature SR spectra of the plain windows (blue dots at top curves) and of two metamaterials (red dots). More spectra are shown in the Supplementary Materials. The black solid curves are the fits using Eqn.1. The dashed blue line is a fit assuming $\text{Im}[C_3]=0$ for the metamaterial at $\lambda_p=858$nm. The residues correspond to the metamaterial at $\lambda_p=858$nm. The units are the same than for the main top curves. A.U., arbitrary units.

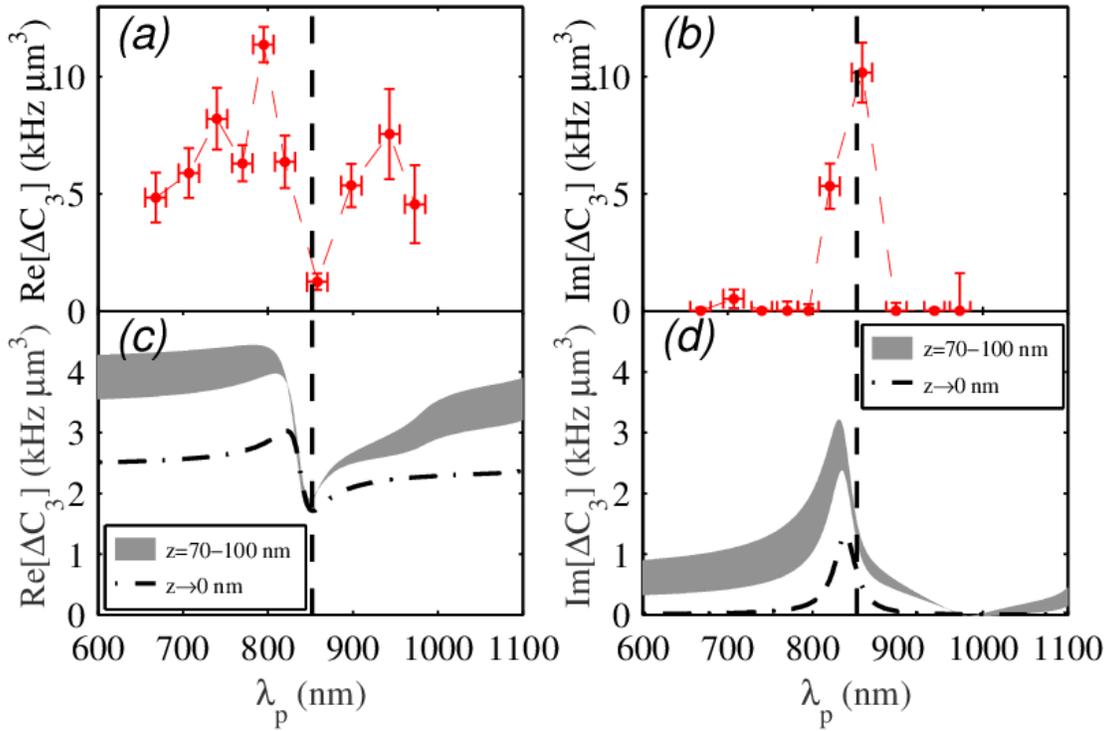

**Fig. 4. The van der Waals coefficient.** ΔC$_3$ coefficients as function of λ$_p$, the position of the plasmon resonance. Real part (a) and imaginary part (b) extracted from the fits of the SR signals. (c) and (d) are the real and imaginary parts of the C$_3$ coefficients computed from the model. The dot-dashed curve corresponds the non-retarded case (z→0). The retarded contribution is taken into account by considering an effective distance ranging from 70 to 100nm. It corresponds to the shaded grey surface. The vertical dashed lines indicate the position of the atomic resonance.

**Supplementary Materials: Tailoring optical metamaterials to tune the atom/surface Casimir-Polder interaction**

**Authors**

Eng Aik Chan,[1,3] Syed Abdullah Aljunid,[1] Giorgio Adamo[1], Athanasios Laliotis[2], Martial Ducloy*[1,2,3], David Wilkowski*[1,3,4,5]

* Corresponding authors' emails: martial.ducloy@univ-paris13.fr, david.wilkowski@ntu.edu.sg

**Affiliations**

[1] Centre for Disruptive Photonic Technologies, TPI, Nanyang Technological University, 637371 Singapore.

[2] Laboratoire de Physique des Lasers, UMR 7538 du CNRS, Université Paris13-Sorbonne-Paris-Cité    F-93430 Villetaneuse, France.

[3] School of Physical and Mathematical Sciences, Nanyang Technological University, 637371, Singapore.

[4] Centre for Quantum Technologies, National University of Singapore, 117543 Singapore.

[5] MajuLab, CNRS-UNS-NUS-NTU, Université Côte d'Azur, International Joint Research Unit UMI 3654, Singapore.

**Stationary surface plasmon waves**

We derive here the general solutions of electromagnetic waves generated by a two dimensional periodic nano-structured metamaterial on one of its main axes. The system is illuminated by a plane wave at normal incidence (see Fig. S1). Since the metamaterial is a thin layer, $l << \lambda$, we will ignore the propagation of the field in the metamaterial and use the Raman-Nath approximation. Hence the tangential wave vectors are $2\pi m/\Lambda$, where $m$ is an integer and $\Lambda$ the period of the metamaterial. Conservation of the frequency imposes

$$k_{out}^2(m) = k_{in}^2 - (2\pi m/\Lambda)^2 \qquad (S1)$$

Here $k_{out}$ and $k_{in}$ are respectively the transmitted wave vector at normal incidence and the incident wave vector. Without any loss of generality in our argumentation, we consider indexes of refraction equal to one. If $m = 0$, the solution is trivially a transmission of the incident plane wave. If $m \neq 0$, since $2\pi/\Lambda > k_{in}$, the normal outgoing vector is negative. Hence the waves are evanescent, decaying exponentially with respect to the normal to the metamaterial. Those solutions are the well-known surface plasmon (SP) waves.

Thus, the general solution can be decomposed into a propagating plane wave at normal incidence and surface plasmon waves. In the case of our two-dimensional metamaterial (see main article for more details), FDFD results are shown on Fig. S1b. The surface plasmon waves are clearly observed if one removes the propagation component as it is shown in Fig. S1c.

Because of the underlying symmetry of the metamaterial, one gets

$$\left|E_{+,m}^{SP}\right| = \left|E_{-,m}^{SP}\right|. \qquad (S2)$$

$|E^{SP}_{+,m}|$ ($|E^{SP}_{-,m}|$) is the amplitude of the surface plasmon wave, with a real tangential wavevector $k^{SP}_{+,m} = 2\pi|m|/\Lambda$ ($k^{SP}_{+,m} = 2\pi|m|/\Lambda$) From Eq. (S2), we find that the surface plasmon waves are stationary waves. It also means that there is no static component corresponding to a null wavevector along the tangential plane.

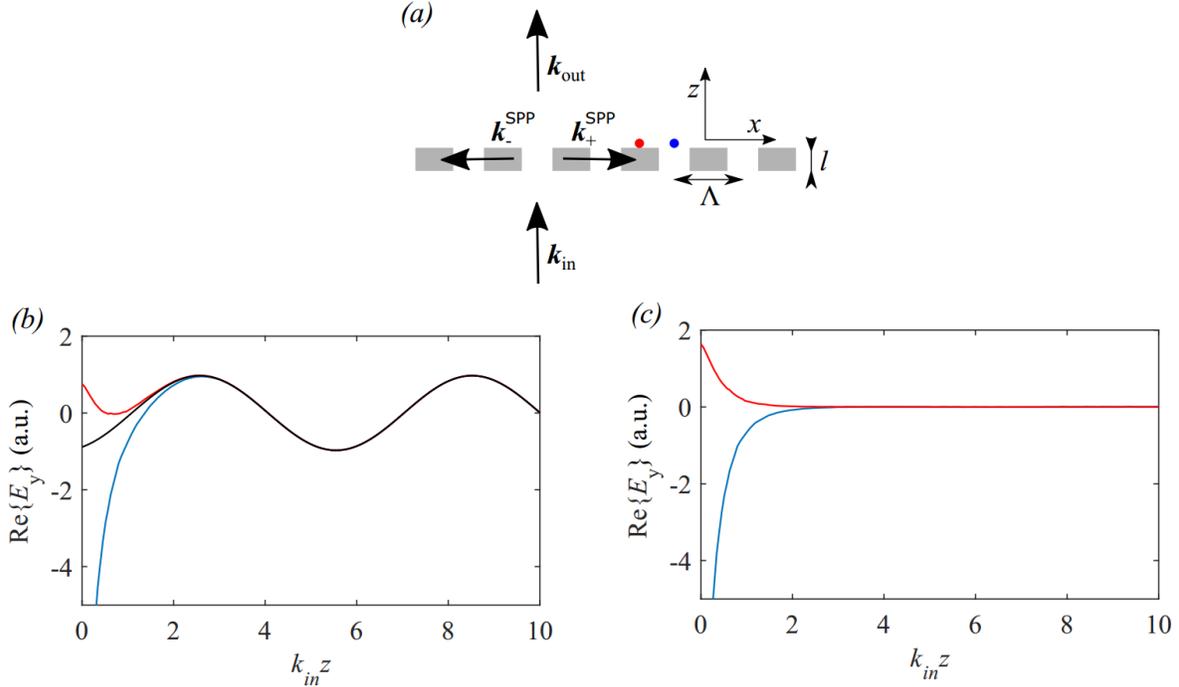

Figure S1. **Electric field simulations** (a) Schematic of the metamaterial and the different wave vectors of the fields (b) Spatial evolution of the real part of the transmitted electric field along *y*, which corresponds to the incident polarization direction. We use here FDFD numerical methods. $z = 0$ correspond to the outer interface of the metamaterial. The blue (red) curves correspond to the center (150 nm away for the center) of the nano-slit. The locations are indicated on (a) by a blue (red) dot. The black curve is a fit of the far-field response. It represents the propagating component of the transmitted field. It is also the transmitted field obtained using the meanfield approximation, *i.e.* replacing the metamaterial by a bulk material having the same far-field properties. (c) same as (b) but the propagating component has been removed. The remaining fields correspond to the surface plasmon wave localized in the vicinity of the metamaterial.

We test this statement on our metamaterial using FDFD simulation and we find that the averaged plasmon field amplitude in the tangential plane at the metamaterial/vacuum interface is small, $i.e. \cong 15\%$ of the propagating field amplitude. It is however not null as predicted by the simple model introduced here. We believe that this mismatch is due either, to the finite thickness of the metamaterial or to some systematic errors in the simulation.

**Mean field approximation for the atomic vapour response**

To analyze the selective reflection spectra, presented in the main article, we modeled the metamaterial by a spatially homogeneous bulk material having the same thickness and the same far-field properties at normal incidence. Then following references [29], [30], we deduce, *n*, the complex index of refraction of the effective bulk material, a mean-field approximation known to well describe the far-field. We check that it is indeed the case in our system and we find an excellent agreement between the bulk and the metamaterial transmitted and reflected fields in the far-field (See Fig. S1b for transmission. Reflection is not shown here).

We note that the effective bulk material does not generate surface plasmon waves and the far-field propagating wave extends up to the material position, *i.e.* $z = 0$. Thus the near-field electromagnetic environments are very different for the metamaterial and the bulk material. Since the reflectance signal of the atomic vapor emerges from the response of the atoms in the vicinity of the metamaterial, the mean-field approximation can be questioned here. However, we find that it is still a good approximation, because of the nature of the plasmon wave. We show, indeed, that the plasmon wave does not have a static component. This latter corresponds to a null wavevector along the tangential plane. The absence of this static component tells us that plasmon waves do not contribute to the Doppler-free signal. Other components of the plasmon wave have a wavevector amplitude $2\pi|m|/\Lambda > |m|k_{in}$. Thus they give strongly Doppler broadened contribution, well smoothed by the relative slow response of the atomic coherence. We run simulations of optical-Bloch equation for an atom flying above the metamaterial surface to confirm this statement. Finally we conclude that, other than modifying the index of refraction, the plasmon wave does not seem to play other roles in our system.

**Experimental data and fit of selective reflection signal for all metamaterials**

The full span of the selective reflection spectrum of atoms across the ten metamaterials is shown in Fig. S2. Signals for plasmon resonance $\lambda_P$, far-detuned from the atomic transition, are considerably smaller than the spectra near coupled resonances because of the low transmission of the metamaterial at $\lambda = 852nm$ (see for example case $\lambda_P = 973nm$ in Fig.S2). From the fits to the data, we extract the complex value of the $C_3$ coefficient shown in Fig. 3 on the main text.

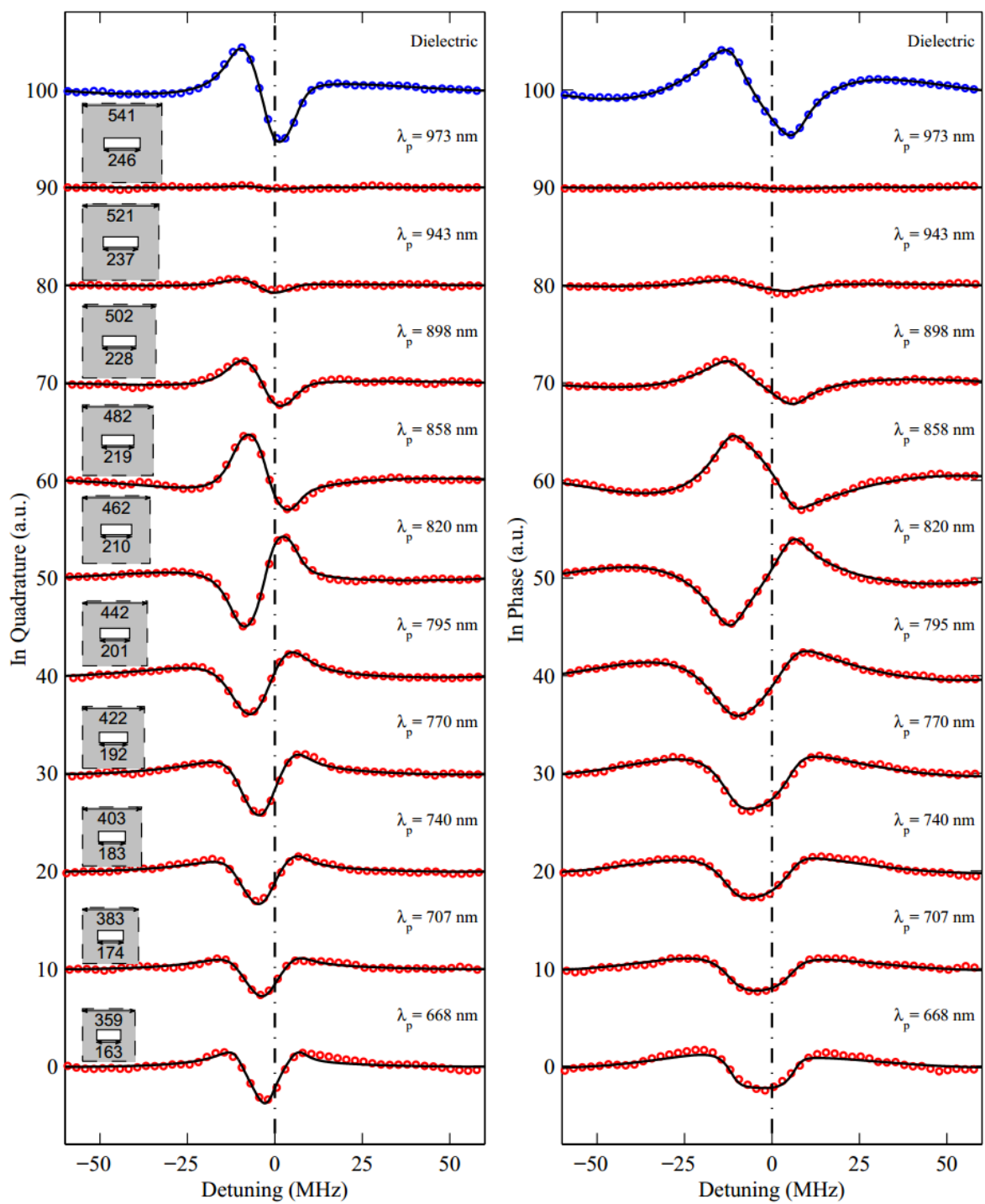

Figure S2. **Selective Reflection spectra** Selective reflection spectra (both in-phase and in quadrature) of the plain windows (blue dots at top curves) and of the ten metamaterials (red dots) with plasmon resonances in main text Fig. 2. Black lines correspond to the fits to the data as explained in the main text.

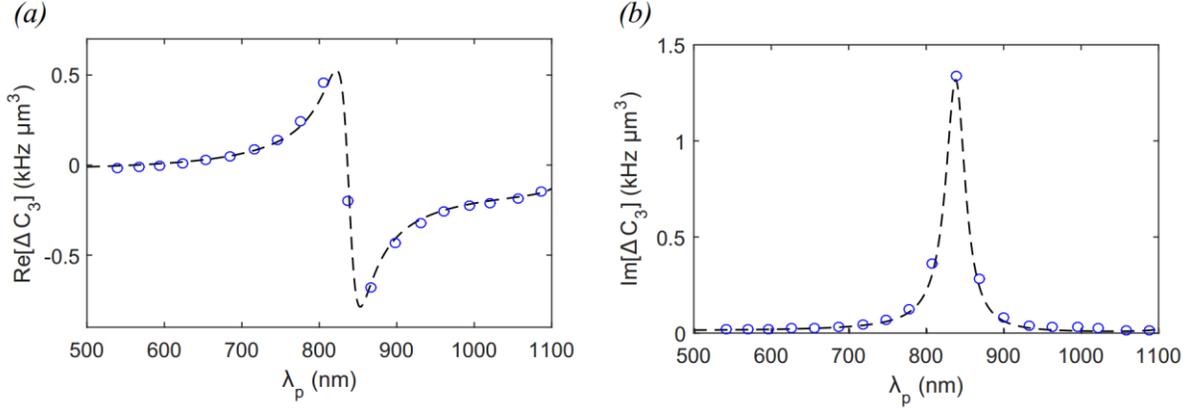

Figure S3. **Resonant contribution to the Casimir-Polder interaction** Comparison between the resonant photon exchange model (blue circle) and the completed model (black dashed line) of
the Casimir-Polder interaction. The plots are as function of the plasmon resonance frequency. The resonant photon exchange model corresponds to Eqn. (S4). (a) Real part of the $C_3$ coefficient, where the background contribution has been removed. (b) Imaginary part of the $C_3$ coefficient.

**Resonant photon contribution to the Casimir-Polder interaction**

An analysis of the Casimir-Polder interaction, in the non-retarded regime, can be done using a simple model where the $C_3$ coefficient is decomposed into two components [33], [34],

$$C_3 = C_3^{\{nr\}} + C_3^{\{r\}}. \quad (S3)$$

Here $C_3^{\{nr\}}$ takes into account the dispersion contribution of virtual exchange of nonresonant vacuum photons, whereas $C_3^{\{r\}}$ gives the contribution due to resonant photon exchange between an excited atom and the metamaterial. The $C_3^{\{nr\}}$ is practically independent of the plasmonic resonance position and is mostly governed by the metallic response over the entire frequency spectrum [33], [34]. The contribution of the plasmon resonance is contained in the $C_3^{\{r\}}$ since it concerns only the resonant photons emitted on the 6P$_{3/2}$ → 6S$_{1/2}$ cesium D2 transition at 852 nm. Following reference [34], we predict,

$$C_3^{\{r\}} = \frac{\mu_r^2}{24\pi\epsilon_0}\frac{\epsilon-1}{\epsilon+1}, \quad (S4)$$

where $\epsilon$ is the complex dielectric constant of the metamaterial extracted from the FDFD simulation and $\mu_r$ the reduced dipole moment of the 6P$_{3/2}$ → 6S$_{1/2}$ coupling. The results of this model are shown in Fig. S3 and compared to the completed approach including non-resonant contribution as discussed in the main article. We see an excellent agreement confirming that the resonance behavior of the Casimir-Polder interaction is due to the resonant photon exchange between an excited atom and the metamaterial.

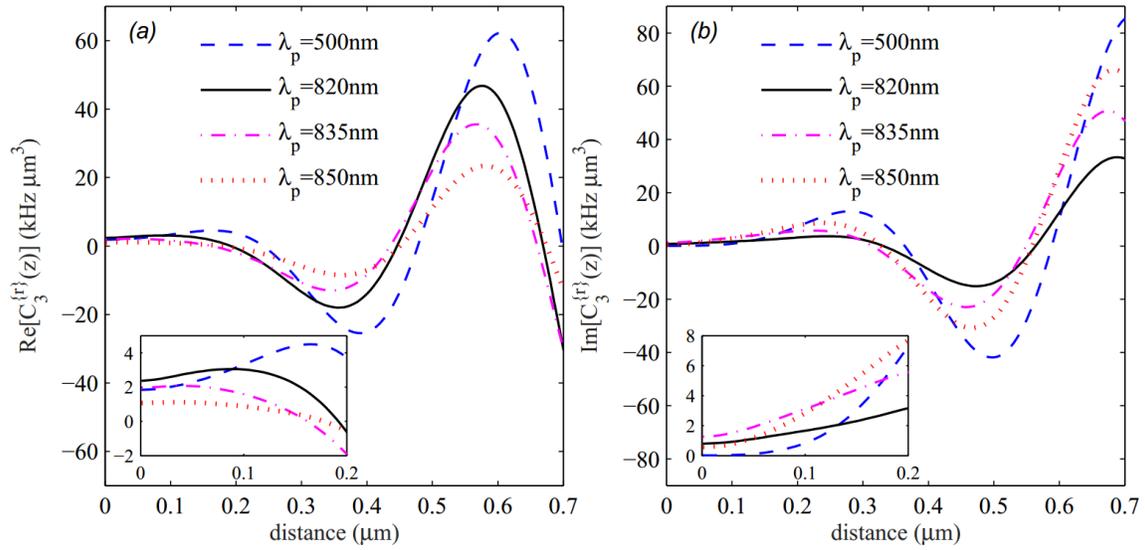

Figure S4. **Retardation effects on selective reflection** Distance dependence of the real (a) and imaginary (b) parts of the resonant component of the effective $C_3(z)$ coefficient for different metamaterials. The values of the plasmon resonances are shown on the figure. We can see that the dispersive shift and the dissipative linewidth oscillate out of phase in the far-field, both with a period of $\lambda/2$. The inset shows a zoom for the first 200 nm, the distance range relevant for selective reflection experiments. QED oscillations have a significant influence on the effective $C_3(z)$ coefficient even in the nanometric range.

**Retardation effects on selective reflection**

As discussed in the previous section, the physics of the coupling between the atom and the surface resonance is contained in the resonant term of the Casimir-Polder interaction that in the near-field ($z \ll \lambda/4\pi$) is given by $\sim C_3^{\{r\}}/z^3$, where $C_3^{\{r\}}$ is a complex and constant coefficient. In the far-field ($z \gg \lambda/4\pi$), $C_3^{\{r\}}$ is not constant anymore. Its dispersive part (shift) asymptotically tends to $\sim |r|\cos(4\pi z/\lambda + \phi)/z$ whereas its dissipative part (linewidth) tends to $|r|\sin(4\pi z/\lambda + \phi)/z$, where $|r|$, and $\phi$ are the modulus and phase of the reflection coefficient. The above considerations suggest that retardation effects cannot be ignored at distances comparable to the probing depth of selective reflection, typically on the order of $\lambda/2\pi$, where the interaction is in an intermediate regime and has no simple analytical form. In Fig.S4 we show the real and imaginary parts of the effective $C_3^{\{r\}}(z)$ coefficient as a function of distance for different metamaterials. The retarded Casimir-Polder interaction decays slowly as a function of distance and displays QED oscillations related to the reflection of spontaneously emitted photons on the meta-surface. The phase and amplitude of these oscillations varies as a function of the plasmonic resonance of the metamaterial. Fig. S4 shows that retardation effects play a significant role in both the amplitude and the shape of the resonant behavior of the $C_3^{\{r\}}(z)$ coefficient. In the main text we investigate a distance range of $70 - 100 nm$, chosen to reside in the heart of the probing depth of selective reflection spectroscopy (a layer of $\lambda/2\pi$ thickness), but also corroborated by a preliminary analysis of selective reflection spectra with a fully retarded Casimir-Polder potential.

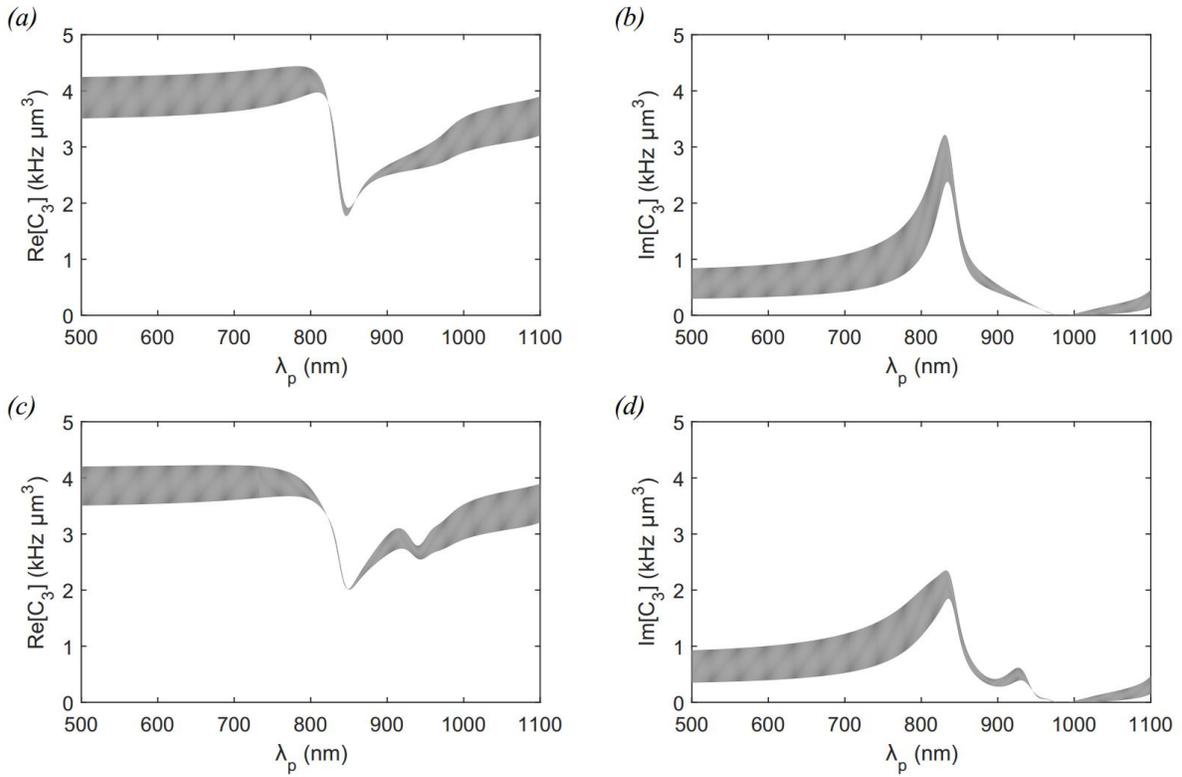

Figure S5. **Finite thickness of the metamaterial** Complete model of the $C_3$ coefficients taking into account the retardation effect in the Casimir Polder interaction (see main article for more details). (a) and (b) are respectively the real and imaginary part of the $C_3$ coefficients considering a semi-infinite effective bulk material for atom distance varying between 70 to 100 nm. (c) and (d) are the same that (a) and (b) but with an effective bulk material having the same thickness with the metamaterial used in our experiment

**Finite thickness of the metamaterial**

The calculation of the $C_3$ coefficients, discussed in the main article, were performed on a semi-infinite effective bulk material (Fig. S5a-b). Here we also investigate the case where the thickness of the bulk material corresponds to the metamaterial used in the experiment (see Fig. S5c-d). Away from the plasmon resonance, the metamaterial strongly absorbs the electromagnetic field. Thus virtual photons and real photons mainly do not reach the end of the metamaterial, i.e. the metamaterial/dielectric interface. We observe very small difference between the semi-infinite and finite cases. Close to the plasmon resonance, the metamaterial becomes more transparent for real photons. The $C_3$ coefficients are slightly affected by the finite size of the material mostly due to the modification of the metamaterial reflectivity (see previous section). This is demonstrated by the appearance of a much smaller extra resonance related to the etalon effect of the metamaterial layer at $\lambda_p \sim 930 nm$. Nevertheless, the modifications remain small, compared to the case of a semi-infinite metamaterial. Most importantly, the etalon effects do not seem to contribute to an increase of the observed amplitude of the $C_3$ resonance.